\documentclass[twocolumn,english,showpacs,preprintnumbers,prb]{revtex4-1}
\usepackage[latin9]{inputenc}
\usepackage{color}
\usepackage{amsmath}
\usepackage{amssymb}
\usepackage{graphicx}
\usepackage{esint}

\makeatletter

\@ifundefined{textcolor}{}{%
 \definecolor{BLACK}{gray}{0}
 \definecolor{WHITE}{gray}{1}
 \definecolor{RED}{rgb}{1,0,0}
 \definecolor{GREEN}{rgb}{0,1,0}
 \definecolor{BLUE}{rgb}{0,0,1}
 \definecolor{CYAN}{cmyk}{1,0,0,0}
 \definecolor{MAGENTA}{cmyk}{0,1,0,0}
 \definecolor{YELLOW}{cmyk}{0,0,1,0}
}

\usepackage{ulem}

\usepackage{aecompl}

\usepackage{epsfig}\usepackage{dcolumn}\usepackage{bm}

\usepackage{babel}

\makeatother

\usepackage{babel}
\begin{document}

\title{Realization of anomalous multiferroicity in free-standing graphene
with magnetic adatoms}

\author{Y. Marques$^{1}$, L. S. Ricco$^{1}$, F. A. Dessotti$^{1}$, R.
S. Machado$^{1}$, I. A. Shelykh$^{2,3,4}$, M. de Souza$^{5}$, and
A. C. Seridonio$^{1,5}$}

\affiliation{$^{1}$Departamento de F\'{i}sica e Qu\'{i}mica, Unesp - Univ Estadual
Paulista, 15385-000, Ilha Solteira, SP, Brazil\\
 $^{2}$Division of Physics and Applied Physics, Nanyang Technological
University 637371, Singapore\\
 $^{3}$Science Institute, University of Iceland, Dunhagi-3, IS-107,
Reykjavik, Iceland\\
 $^{4}$ITMO University, St. Petersburg 197101, Russia\\
 $^{5}$IGCE, Unesp - Univ Estadual Paulista, Departamento de F\'{i}sica,
13506-900, Rio Claro, SP, Brazil}
\begin{abstract}
It is generally believed that free-standing graphene does not demonstrate
any ferroic properties. In the present work we revise this statement
and show that single graphene sheet with a pair of magnetic adatoms can be driven into ferroelectric (FE) and multiferroic
(MF) phases by tuning the Dirac cones slope. The transition into the
FE phase occurs gradually, but an anomalous MF phase appears abruptly
by means of a Quantum Phase Transition. Our findings suggest that
such features should exist in graphene recently investigated by Scanning
Tunneling Microscopy (Science \textbf{352}, 437 (2016)).
\end{abstract}


\maketitle

\section{Introduction}

Multiferroic (MF) materials, i.e., compounds where ferroelectricity
and magnetic ordering coexist, have been recognized as systems with
large applications in the modern device industry\cite{Khomskii,Eerenstein}.
Nowadays, it is well-known that the mechanisms behind the multiferroic
behavior are not universal and thus often material specific\cite{Review1,Review2}.
The origin of the MF response is thus a puzzling issue which attracts
the attention of researchers working in the domains of both condensed
matter physics and materials science. Prominent examples where multiferroicity
can be found include frustrated magnets\cite{Kimura,Hur,Cheong};
systems involving the Dzyaloshinskii-Moriya interaction as observed
in manganites with spiral spin-order\cite{Fiebig,Cheong}, in which
the concomitant formation of electric-dipoles and long-range magnetic
ordering takes place; the recently reported 2D systems with ``Mexican-hat''
type band-structure\cite{MexicanH}; incommensurate states with broken
lattice inversion symmetry\cite{FeTe}; and molecular conductors,
where the MF behavior is linked to electron-electron correlations\cite{FSalts,Mariano}.

\begin{figure}
\includegraphics[height=0.25\textheight]{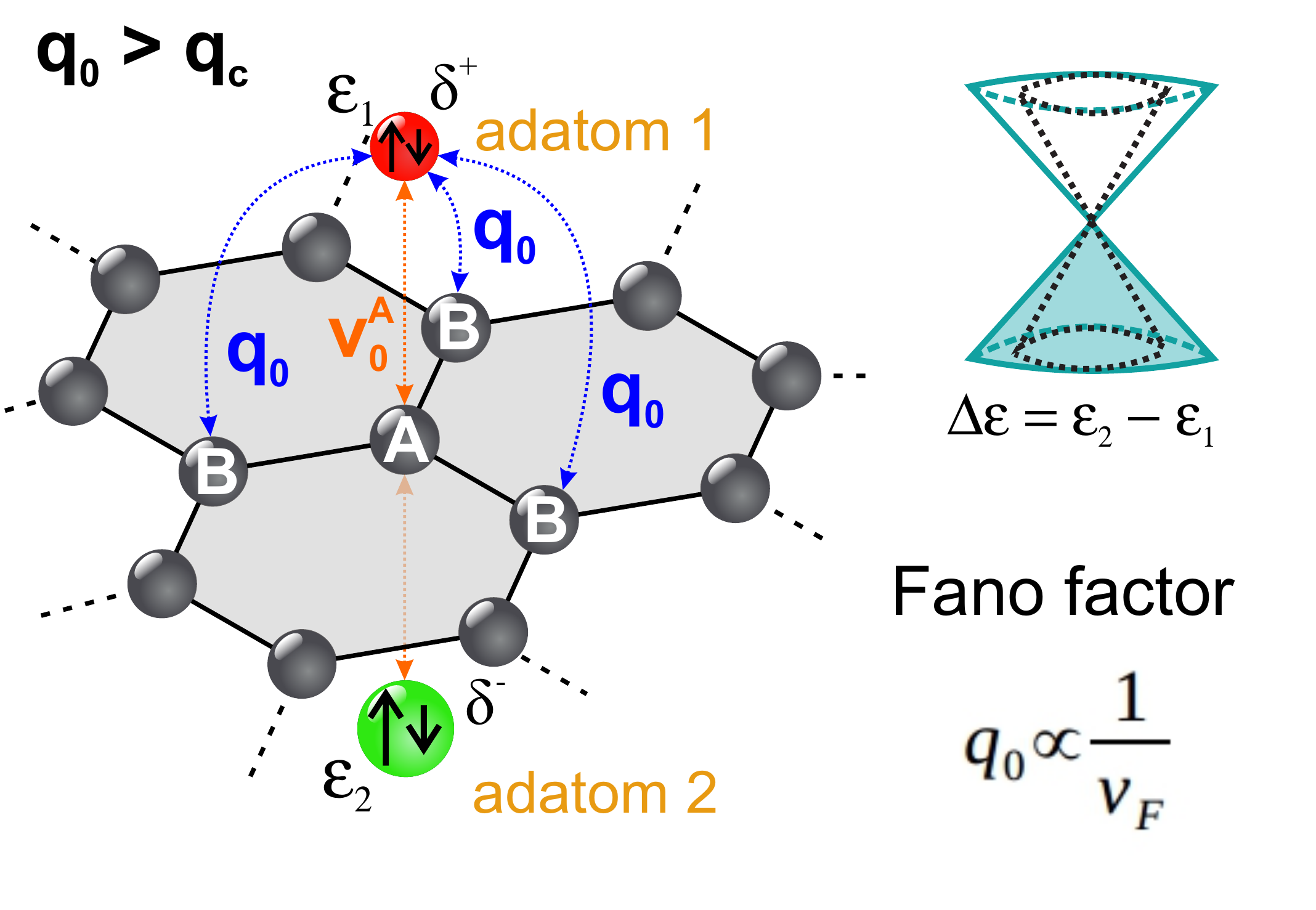}
\caption{\label{fig:Pic1} (Color online) MF phase of two energetically different
magnetic adatoms ($\Delta\mathcal{E}$) collinear
to a carbon: distinct charge accumulations $\delta^{+}$ and $\delta^{-}$
together with a net magnetization (vertical arrows) can split over
these adatoms. In this system, a QPT modifies abruptly the FE phase
into the MF, due to the increasing of the Fano factor $q_{0}>q_{c}$
above the critical point, via the tuning of the Dirac cones slope
(the Fermi velocity $v_{F}$).}
\end{figure}

In this work, we investigate the perspective of using single graphene
monolayer for achievement of MF behavior. It is well known that free-standing
graphene is not suitable for this purpose\cite{Dresselhaus,CNeto,BUchoa},
thus we propose to add a pair of collinear magnetic
adatoms to it situated from different sides of the sheet as shown at Fig.\ref{fig:Pic1}. The onset of the MF phase is achieved
by tuning the slope of the Dirac cones, which breaks adatom-graphene
sublattice symmetry at some critical point and drives the system towards
a Quantum Phase Transition (QPT) from the ferroelectric (FE) phase
to the MF.

This multiferroicity is anomalous, once
this phase is preceded by a purely FE phase just with charge ordering
where the local magnetic moments of the adatoms are quenched, in opposite
to the conventional MF behavior, characterized by ordering parameters
of ferroelectric and magnetic-type emerging concomitantly\cite{Review1,Review2}
and also due to the lack of conjugate fields (electric and magnetic)
as tuning parameters of the phase transition. Here, we show that both
can be replaced by just a single tuning parameter, which is established
by the Fano factor $q_{0}$ of interference responsible for changing
the Dirac cones slope. In this way, standard hysteresis loops depending upon the conjugate fields for identifying ferroelectricity and the single-phase multiferroics are not required within our framework.

\section{The model}

To give theoretical description of the system under consideration,
the Anderson-type Hamiltonian\cite{BUchoa,Anderson} can be proposed:
\begin{align}
\mathcal{H}^{\text{2D}} & =-t\sum_{\bold k\sigma}[\phi({\bold k})a_{\bold k\sigma}^{\dagger}b_{\bold k\sigma}+\text{H.c.}]+\sum_{l\sigma}\mathcal{E}_{l}d_{l\sigma}^{\dagger}d_{l\sigma}\nonumber \\
 & +[\mathcal{V}_{0}^{A}\sum_{\bold kl\sigma}(a_{\bold k\sigma}^{\dagger}+\frac{t}{D}q_{0}\phi({\bold k})b_{\bold k\sigma}^{\dagger})d_{l\sigma}+\text{H.c.}]\nonumber \\
 & +\sum_{l}\mathcal{U}n_{l\uparrow}n_{l\downarrow.}\label{eq:TIAM}
\end{align}
In this expression, $D$ is the bandwidth, $\phi({\bold k})=\sum_{i=1}^{3}e^{i{\bold k}\cdot{\bold\delta}_{i}},$
${\bold\delta_{1}}=a{\bold e}_{x}$ and ${\bold\delta}_{2,3}=\frac{a}{2}(-{\bold e}_{x}\pm\sqrt{3}{\bold e}_{y})$
are the nearest neighbor vectors and $a$ is the side length of the
hexagonal cell. The surface electrons are described by the operators
$a_{\bold k\sigma}^{\dagger}$ ($a_{\bold k\sigma}$) and $b_{\bold k\sigma}^{\dagger}$
($b_{\bold k\sigma}$) for the creation (annihilation) with momentum
${\bold k}$ and spin $\sigma,$ respectively in the sublattices\textit{
$A$} and \textit{$B$}. For the adatoms, $d_{l\sigma}^{\dagger}$
($d_{l\sigma}$) creates (annihilates) an electron with spin $\sigma$
in the state $\mathcal{E}_{l},$ wherein $l=1,2$. The third term
in the expression \ref{eq:TIAM} mixes the continuum of the graphene
states with localized levels of the adatoms $\mathcal{E}_{l}.$ This
hybridization is described by the local tunneling term $\mathcal{V}_{0}^{A}$
corresponding to the electron hopping between a carbon atom and a
pair of adatoms flanking it, as it is shown at Fig.\,\ref{fig:Pic1}.
The Fano factor\cite{Fano1}{} is the control parameter
of the QPT in our system and is given by the following ratio:
\begin{equation}
q_{0}=\frac{v}{v_{F}},\label{eq:Fanoq}
\end{equation}
which is proportional to the constant $v=\frac{3aD}{2\hbar}\frac{\mathcal{V}_{0}^{B}}{\mathcal{V}_{0}^{A}}$
dependent upon $\mathcal{V}_{0}^{A}$ and the couplings $\mathcal{V}_{0}^{B}$
with next three nearest carbon atoms, wherein $\mathcal{V}_{0}^{\alpha}=\int d\bold r[\phi^{\alpha}(\bold r)]^{*}h\phi_{L}(\bold r)$
is a Slater-type bond in the \textit{Linear Combination
of Atomic Orbitals} approach expressed in terms
of the $\pi$ orbitals $\phi^{\alpha}(\bold r)$ for the sublattice
$\alpha,$ the localized adatom wavefunction $\phi_{L}(\bold r)$
and the single-particle Hamiltonian $h$, also inversely proportional
to the Fermi velocity $v_{F}=\frac{3at}{2\hbar}$\cite{Overlaps}.

Regarding the chemical bond between the carbon atom
at the sublattice A and the two collinear magnetic adatoms depicted
in Fig.\ref{fig:Pic1}, we expect that the metallic-type
bond should stabilize the set of adatoms in graphene. However, the
accuracy of our model should be verified by means of an ab-initio
analysis. The latter does not belong to the scope of this work. Our
theoretical framework focus on evaluating the system band-structure
by means of the standard tight-binding method, which for graphene,
takes into account the $\pi$ and $\pi^{*}$ bands formed by $p_{z}$
orbitals placed on carbon atoms in the presence of localized wavefunctions
$\phi_{L}(\bold r)$ for the adatoms. In such a
scenario, solely electronic hopping terms are accounted for the Hamiltonian.
We should pay special attention to the regime $q_{0}>q_{c}$ (the
critical point defined later in the text), once it mimics the strong
coupling limit between the adatoms and the sublattice \textit{$B.$}
In this regime, in particular, adatom-graphene sublattice symmetry
breaking occurs strongly. As we will discuss later on, it is of capital
importance for triggering the QPT and rising of the MF phase. The
last term accounts for the on-site Coulomb interaction $\mathcal{U}$,
with $n_{l\sigma}=d_{l\sigma}^{\dagger}d_{l\sigma}$.

The parameter which characterizes the QPT that can be accessed experimentally
via Scanning Tunneling Microscopy (STM)\cite{STM} is the density
of states (DOS) of the $l^{th}$ adatom:
\begin{equation}
\text{DOS}_{ll}^{\sigma}=-\frac{1}{\pi}{\tt Im}(\tilde{\mathcal{G}}_{d_{l\sigma}d_{l\sigma}}),
\end{equation}
where the Green's function in energy domain $\tilde{\mathcal{G}}_{d_{l\sigma}d_{j\sigma}}$
is a Fourier transform of corresponding function in time domain,
\begin{equation}
\mathcal{G}_{d_{l\sigma}d_{j\sigma}}=-\frac{i}{\hbar}\theta\left(\tau\right){\tt Tr}\{\varrho_{\text{2D}}[d_{l\sigma}\left(\tau\right),d_{j\sigma}^{\dagger}\left(0\right)]_{+}\},
\end{equation}
wherein $\theta\left(\tau\right)$ is the Heaviside function and $\varrho_{\text{2D}}$
is the density matrix of Eq.(\ref{eq:TIAM}). To get expression for
$\tilde{\mathcal{G}}_{d_{l\sigma}d_{j\sigma}},$ we apply the equation-of-motion
(EOM) method using Hubbard I approximation\cite{Hubbard}. In energy
domain, one gets:
\begin{align}
(\mathcal{E}^{+}-\mathcal{E}{}_{l})\tilde{\mathcal{G}}_{d_{l\sigma}d_{j\sigma}} & =\delta_{lj}+\Sigma\sum_{\tilde{l}}\tilde{\mathcal{G}}_{d_{\tilde{l}\sigma}d_{j\sigma}}+\mathcal{U}\tilde{\mathcal{G}}_{d_{l\sigma}n_{l\bar{\sigma}},d_{j\sigma}},\nonumber \\
\label{eq:s1}
\end{align}
with $\mathcal{E}^{+}=\mathcal{E}+i0^{+}$ and
\begin{equation}
\Sigma=(\mathcal{V}_{0}^{A})^{2}\sum_{\bold k}\frac{\mathcal{E}^{+}(1+\frac{t^{2}}{D^{2}}q_{0}^{2}\left|\phi(\mathbf{k})\right|^{2})-2\frac{t^{2}}{D}q_{0}\left|\phi(\mathbf{k})\right|^{2}}{\mathcal{E}^{+2}-t^{2}|\phi({\bold k})|^{2}}\label{eq:SFE}
\end{equation}
is the non-interacting self-energy. Noteworthy, as the magnetic adatoms break the translational invariance of the lattice and time-reversal symmetry, the graphene band structure is affected: by looking at the Anderson broadening\cite{Anderson} $\Delta=-\text{Im}\Sigma=\pi(\mathcal{V}_{0}^{A})^{2}\text{\ensuremath{\mathcal{D}}}_{0}$ with $\text{\ensuremath{\mathcal{D}}}_{0}=\frac{\left|\varepsilon\right|}{D^{2}}(1-q_{0}\frac{\varepsilon}{D})^{2},$ we notice that the new graphene local density of states $\text{\ensuremath{\mathcal{D}}}_{0},$ nearby the Dirac cones, becomes Fano factor dependent as a result.

In the equation above, $\tilde{\mathcal{G}}_{d_{l\sigma}n_{l\bar{\sigma}},d_{j\sigma}}$
provides a two particle Green's function determined by the Fourier
transform of
\begin{equation}
\mathcal{G}_{d_{l\sigma}n_{l\bar{\sigma}},d_{j\sigma}}=-\frac{i}{\hbar}\theta\left(\tau\right){\tt Tr}\{\varrho_{\text{2D}}[d_{l\sigma}\left(\tau\right)n_{l\bar{\sigma}}\left(\tau\right),d_{j\sigma}^{\dagger}\left(0\right)]_{+}\},
\end{equation}
where $\bar{\sigma}=-\sigma$ and $n_{l\bar{\sigma}}=d_{l\bar{\sigma}}^{\dagger}d_{l\bar{\sigma}}.$
In order to close the system of the equations for Green's functions,
we write the expression for $\tilde{\mathcal{G}}_{d_{l\sigma}n_{l\bar{\sigma}},d_{j\sigma}},$
which reads:
\begin{align}
(\mathcal{E}^{+}-\mathcal{E}_{l}-\mathcal{U})\tilde{\mathcal{G}}_{d_{l\sigma}n_{l\bar{\sigma}},d_{j\sigma}} & =\delta_{lj}<n_{l\bar{\sigma}}>+\mathcal{V}_{0}^{A}\nonumber \\
\times\sum_{\bold ks}[-\phi_{s}({\bold k})\tilde{\mathcal{G}}_{c_{s\bold k\bar{\sigma}}^{\dagger}d_{l\bar{\sigma}}d_{l\sigma},d_{j\sigma}} & +\phi_{s}^{*}({\bold k})(\tilde{\mathcal{G}}_{c_{s\bold k\sigma}d_{l\bar{\sigma}}^{\dagger}d_{l\bar{\sigma}},d_{j\sigma}}\nonumber \\
+\tilde{\mathcal{G}}_{d_{l\bar{\sigma}}^{\dagger}c_{s\bold k\bar{\sigma}}d_{l\sigma},d_{j\sigma}}) & ],\label{eq:H_GF_2}
\end{align}
wherein the index $s=A,B$ marks a sublattice, $c_{A\bold k\sigma}=a_{\bold k\sigma}$
and $c_{B\bold k\sigma}=b_{\bold k\sigma}$, $\phi_{A}({\bold k})=1$
and $\phi_{B}({\bold k})=\frac{t}{D}q_{0}\phi({\bold k})$, expressed
in terms of new Green's functions of the same order of $\tilde{\mathcal{G}}_{d_{l\sigma}n_{l\bar{\sigma}},d_{j\sigma}}$
and the occupation number
\begin{equation}
<n_{l\bar{\sigma}}>=\int_{-D}^{+D}n_{F}(\mathcal{E})\text{DOS}_{ll}^{\bar{\sigma}}d\mathcal{E},\label{eq:occupation}
\end{equation}
with $n_{F}(\mathcal{E})$ as the Fermi-Dirac distribution. We decouple
the Green's functions in the right-hand side of Eq.(\ref{eq:H_GF_2})
by employing the Hubbard I approximation, considering $\tilde{\mathcal{G}}_{c_{s\bold k\bar{\sigma}}^{\dagger}d_{l\bar{\sigma}}d_{l\sigma},d_{j\sigma}}$
and $\tilde{\mathcal{G}}_{d_{l\bar{\sigma}}^{\dagger}c_{s\bold k\bar{\sigma}}d_{l\sigma},d_{j\sigma}}$
according to $\tilde{\mathcal{G}}_{\mathcal{A}^{\dagger}\mathcal{B}\mathcal{C},\mathcal{D}}\simeq<\mathcal{A}^{\dagger}\mathcal{B}>\tilde{\mathcal{G}}_{\mathcal{C}\mathcal{D}}.$
$\tilde{\mathcal{G}}_{c_{s\bold k\sigma}d_{l\bar{\sigma}}^{\dagger}d_{l\bar{\sigma}},d_{j\sigma}}$
is obtained via EOM and truncated as previously, which yields
\begin{equation}
\tilde{\mathcal{G}}_{d_{l\sigma}d_{l\sigma}}=\frac{\lambda_{l}^{\bar{\sigma}}}{\mathcal{E}-\mathcal{E}_{l}-{{\tilde{\Sigma}}^{\bar{\sigma}}}_{l\bar{l}}},
\end{equation}
where $\bar{l}=1,2$, $l=2,1$ are indices of the distinct adatoms,
\begin{equation}
{{\tilde{\Sigma}}^{\bar{\sigma}}}_{l\bar{l}}=\Sigma+\frac{\lambda_{l}^{\bar{\sigma}}\lambda_{\bar{l}}^{\bar{\sigma}}\Sigma^{2}}{\mathcal{E}-\mathcal{E}_{\bar{l}}-\Sigma}\label{self}
\end{equation}
is the total self-energy and $\lambda_{l}^{\bar{\sigma}}=1+\mathcal{U}<n_{l\bar{\sigma}}>(\mathcal{E}-\mathcal{E}_{l}-\mathcal{U}-\Sigma)^{-1}$
is the spin-dependent spectral weight. For ferromagnetic (FM) and
MF solutions $\lambda_{l}^{\bar{\sigma}}\neq\lambda_{l}^{\sigma},$
otherwise we have the normal (N) phase or FE.

\section{Results and Discussion}

We model magnetic adatoms by starting with $<n_{l\bar{\sigma}}>\neq<n_{l\sigma}>$
in the self-consistent evaluation of Eq.(\ref{eq:occupation}) and
considering the following relation between the parameters: $\mathcal{V}_{0}^{A}=\mathcal{U}=0.25D$\cite{BUchoa}
and temperature $T=0.$ Noteworthy, from the experimental perspective, our findings
are kept robust solely within the range of extremely low temperatures (mK order), since the phenomenon reported here is triggered by a QPT\cite{Tfinite}.
\begin{figure}
\includegraphics[width=0.5\textwidth,height=0.45\textheight]{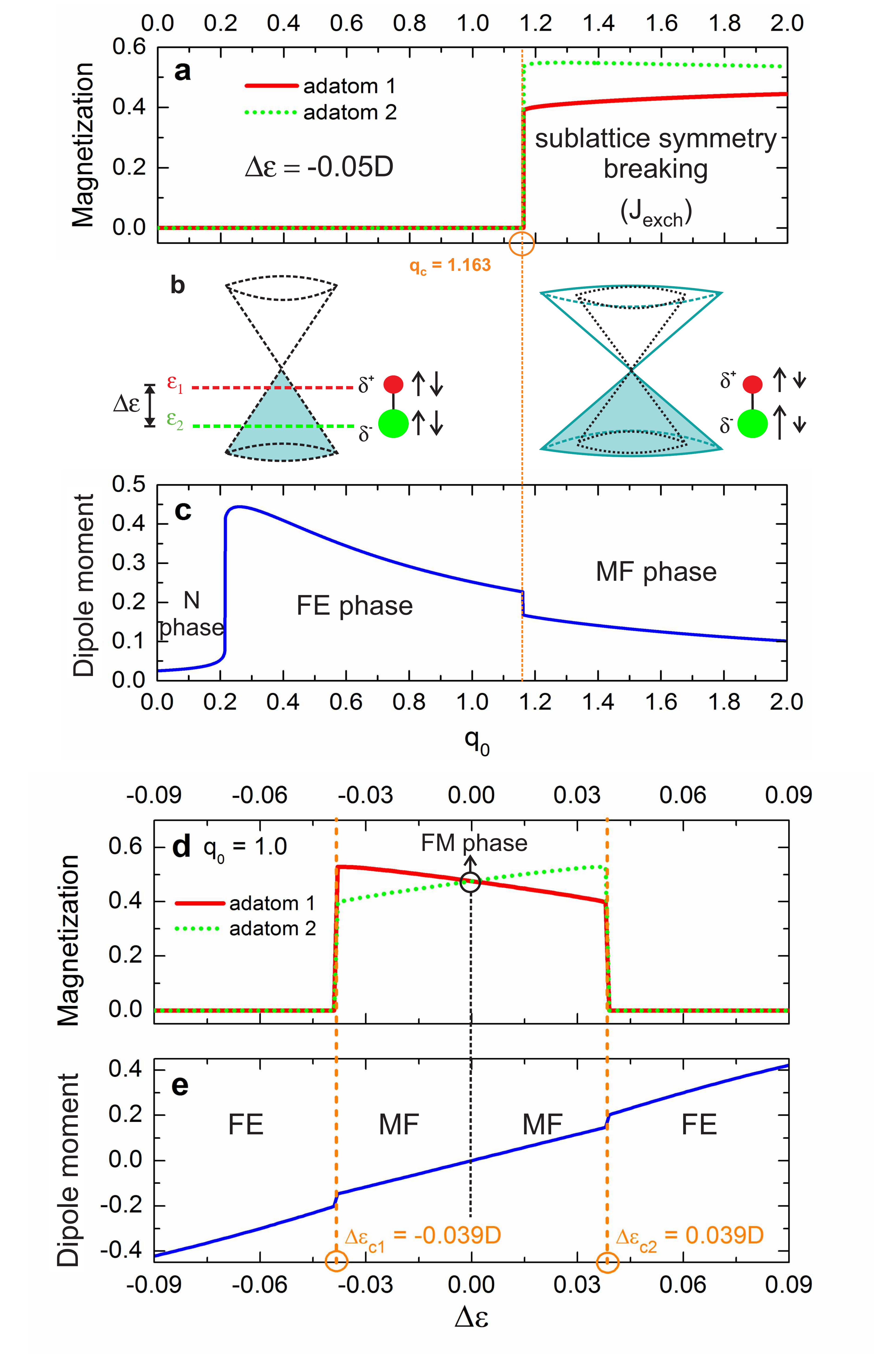}
\caption{\label{fig:Pic2} (Color online) (a) Magnetization of the system as
a function of the Fano factor $q_{0}$ for fixed detuning $\Delta\varepsilon$.
At critical value of the parameter $q_{c}$ the system is driven to
the MF phase by a QPT. (b) The sketch illustrating the connection
between the slope of the Dirac cones and magnetization of the system.
(c) Dipole moment of the system as a function of the Fano factor $q_{0}$
for fixed detuning $\Delta\varepsilon$. After slow initial increase,
the dipole moment experiences fast growing, followed by slow decrease
and discontinuity at $q_{0}=q_{c}$. Rising of an anomalous MF phase
for $q_{0}>q_{c}.$ (d) Magnetization of the system as a function
of the detuning $\Delta\varepsilon$ for fixed Fano factor $q_{0}$.
(e) Dipole moment of the system as a function of the detuning $\Delta\varepsilon$
for fixed $q_{0}$. The MF phase exists within the critical region
$\Delta\varepsilon_{\text{{c}}1}<\Delta\varepsilon<\Delta\varepsilon_{\text{{c}2}},$
except for the FM phase where $\Delta\varepsilon=0.$ Otherwise, just
the FE phase is present.}
\end{figure}
In Fig.\ref{fig:Pic2}, we show the analysis of two spin-degenerate
resonant states nearby and below the Dirac point. We consider the
case for which the detuning between the energies of the two adatoms
is non-zero, $\Delta\mathcal{E}=\mathcal{E}_{2}-\mathcal{E}_{1}=-0.05D,$
which leads to the appearance of the dipole moment $\delta^{-}-\delta^{+}=(<n_{2\uparrow}>+<n_{2\downarrow}>)-(<n_{1\uparrow}>+<n_{1\downarrow}>).$

For the Fano factor lying within the range $0.2<q_{0}<q_{c}=1.163$
(the critical point), the system is characterized only by the FE phase,
since the MF behavior is absent as magnetizations $m_{l}=<n_{l\uparrow}>-<n_{l\downarrow}>$
of the adatoms are zero as can be seen at panels (a) and (b) of the
same figure, where we verify that the local magnetic moments of the
adatoms do not survive when embedded into graphene system. Fig.\ref{fig:Pic2}(c)
illustrates the dipole moment of the pair of adatoms. We
should stress that $\delta^{-}-\delta^{+}$ and $m_{l}$ are local
order parameters at the collinear sites of the adatoms, respectively
for electric and magnetic degrees of freedom. Generally, the FE feature
(electric polarization) of a system can be well-marked by evaluating
the Berry phase\cite{Review1,book}, which depends upon the delocalized
electronic Wannier functions spread over the crystal lattice. Here
such an approach can not be invoked, since $\delta^{-}-\delta^{+}$
arises exclusively from the local FE feature of the adatoms, which
are expected to exhibit wavefunctions extremely localized at their
sites. As a result, this characteristic then prevents the use of the
Berry phase method to recognize a phase as FE with long-range order
parameter. In the range $q_{0}<0.2,$ $\delta^{-}-\delta^{+}$ is
almost null as expected for the N phase, and then increases rapidly
signifying the formation of the FE phase.

Physically, the FE feature, which here is not spontaneous, appears due to the charge
imbalance between the magnetic adatoms, which is a purely electronic
effect caused by the non-zero detuning $\Delta\mathcal{E}$ of their
energy levels and the natural narrowing of the Dirac cones in the
band structure outlined in Fig.\ref{fig:Pic2}(b): the closer to the
Dirac point is the adatom energy level $\mathcal{E}_{1}$ embedded
in the graphene band structure, more emptied of electrons it should
be (smaller red sphere with charge accumulation $\delta^{+}$ ). While
the deeper is the embedded level $\mathcal{E}_{2}$ below the Dirac
point, more electrons will accumulate in this adatom (bigger green
sphere with charge accumulation $\delta^{-}$ ). Thus, as the graphene
density of states decreases if one moves to the Dirac point (intrinsic
bottleneck shape of the Dirac cones), the quantities $\delta^{+}$
and $\delta^{-},$ respectively for the levels $\mathcal{E}_{1}$
and \textit{$\mathcal{E}_{2}$} will be different as a result.

It is worth mentioning that in our approach, the
formation of electric dipoles is thereby purely of electronic origin
as we have discussed above. Lattice distortion is a direct consequence
of the formation of such. A similar situation can be found in molecular compounds at the Mott metal-to-insulator\cite{Mariano1,Mariano2,Mariano3}
and charge-ordering transitions\cite{Mariano4,Mariano5}.
Although our approach does not cover lattice effects
(ferroelasticity), given the presence of adatoms above and below the
carbon atom (see Fig.\ref{fig:Pic1}), one should expect that the sublattices move out-of-plane, but in opposite directions due to the charge imbalance $\delta^{+}$ and $\delta^{-}$ giving rise to a local lattice distortion. The evaluation of the electric dipole here concerns the sites of the adatoms and not a net contribution, being our analysis unaffected by the lattice distortion. To know the entire response, an ab-initio analysis should be implemented in order to find out if the polar catastrophe occurs compensating the electric dipole due to the adatoms. Here, we focus just on such a contribution.

Concerning the MF phase, our findings demonstrate that magnetic ordering
$m_{l}$ is formed abruptly by means of a QPT when
the Fano factor reaches its critical value $q_{0}=q_{c}$ as seen
at Fig.\ref{fig:Pic2}(a). At the same time, the value of the dipole
moment $\delta^{-}-\delta^{+}$ experiences a jump
down at $q_{0}=q_{c}$ (see Fig.\ref{fig:Pic2}(c)). However, its
value still remains non-zero and thus the system reaches the MF phase.
We emphasize that both local order parameters $\delta^{-}-\delta^{+}$
and $m_{l}$ are finite and become coupled to each other (magnetoelectric
effect) just for $q_{0}\geq q_{c},$ once for $q_{0}<q_{c}$ only
$\delta^{-}-\delta^{+}\neq0$ exists and local magnetic moments of
adatoms are completely suppressed. This way, distinctly from standard
multiferroicity, conjugate fields as electric and magnetic are not
required for connecting these order parameters, being the Fano factor
the unique control (tuning) parameter responsible for establishing
the aforementioned correlation and the QPT as well. As a result, such
features characterize the MF behavior here reported as anomalous. However, we do not discard that the conjugate fields (electric and magnetic) can change simultaneously the charge and magnetic order parameters, since such fields be applied to the system in the regime $q_{0}>q_{c},$ when the order parameters become correlated. For this situation, the multiferroicity here addressed would be ruled by the conjugate fields as usually occurs for bulk single-phase multiferroics.

As the Fano factor is inversely proportional to Fermi velocity (see
Eq.(\ref{eq:Fanoq})), the latter can be used as tuning parameter
driving the QPT. Notice that in the MF phase, adatoms magnetize with the same sign revealing that they exhibit parallel magnetic moments, which means that emerging effective exchange coupling
of their spins $J_{\text{{exch}}}=\text{{Re}}[\lambda_{l}^{\bar{\sigma}}\lambda_{\bar{l}}^{\bar{\sigma}}\Sigma^{2}(\mathcal{E}-\mathcal{E}_{\bar{l}}-\Sigma)^{-1}]$
(see Eq.(\ref{self})) is of Ruderman-Kittel-Kasuya-Yosida ferromagnetic-type, which manifests via graphene host as the self-energy $\Sigma$ ensures. Such a coupling arises
from the increasing of the Fano factor value, which breaks strongly
the adatom-graphene sublattice symmetry above the critical point,
since the terms $\frac{t}{D}q_{0}\phi({\bold k})b_{\bold k\sigma}^{\dagger}d_{l\sigma}+\text{{H.c.}}$
become more pronounced with respect to $a_{\bold k\sigma}^{\dagger}d_{l\sigma}+\text{{H.c.}}$
in Eq.(\ref{eq:TIAM}) for this regime. As aftermath,
we find magnetic solutions and $J_{\text{{exch}}}$ is turned-on abruptly
yielding the MF phase. This result matches the experimental findings
reported in Ref.{[}\onlinecite{STM}{]}, where a ferromagnetic coupling
between magnetic adatoms is established by an exchange interaction via graphene,
which is peculiar in a such a system: for the scenario of magnetic
adatoms placed at carbon atoms belonging to the same sublattice, their
local magnetic moments persist within the graphene environment and
$J_{\text{{exch}}}$ is ferromagnetic-type, otherwise these moments
become suppressed.

Here, the former situation occurs for $q_{0}\geq q_{c}$
and once in this range the Fano factor $q_{0}$ attains higher values,
it forces the adatoms to perceive solely the sublattice B leading
to $J_{\text{{exch}}}$ of ferromagnetic-type, while below the critical
point $q_{0}<q_{c}$ the strength of $q_{0}$ is moderate, thus turning-off
the magnetism at the adatoms, as aftermath of their couplings at the
same footing with both sublattices. This situation should be distinguished with respect to the hollow setup
considered by some of us\cite{LH}, where a ferromagnetic exchange is not verified even considering magnetic adatoms. In such a case, the
magnetic moments of the adatoms, within the Hubbard I approximation, become quenched as pointed out by our self-consistent calculations. For the bridge configuration, which is similar to the hollow case, we let it for the near future.
\begin{figure}
\includegraphics[width=0.5\textwidth,height=0.37\textheight]{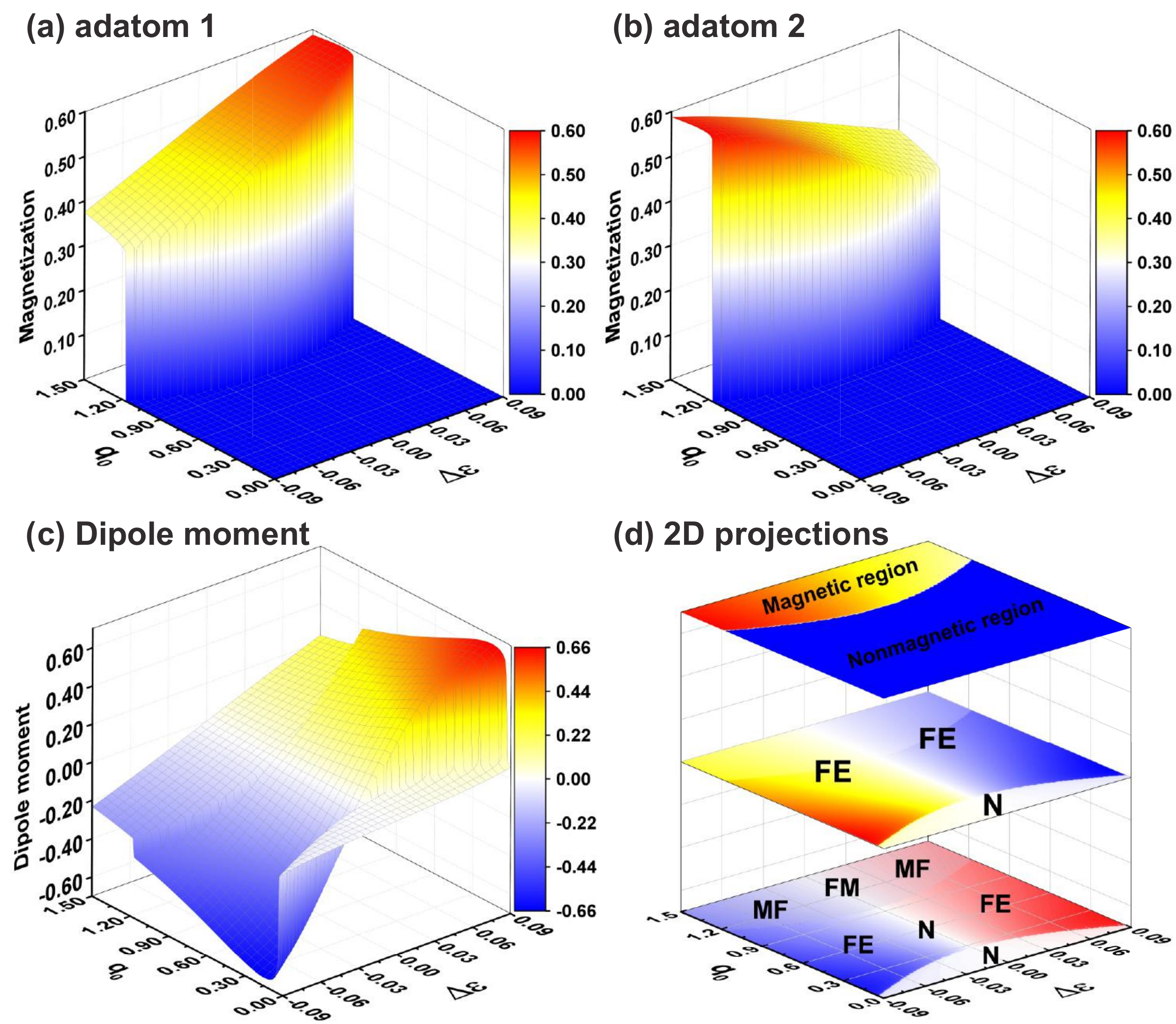}
\caption{\label{fig:Pic3}(Color online) (a)-(b) Magnetizations of the adatoms
as a function of both $q_{0}$ and $\Delta\varepsilon$. (c) The dipole
moment of an adatom pair as a function of both $q_{0}$ and $\Delta\varepsilon$
and finally (d), Upper plot: phase diagram for FM order parameter,
middle plot: phase diagram for FE order parameter and lower plot:
total phase diagram for N, FE, FM and MF phases.}
\end{figure}

\begin{figure}
\includegraphics[width=0.5\textwidth,height=0.32\textheight]{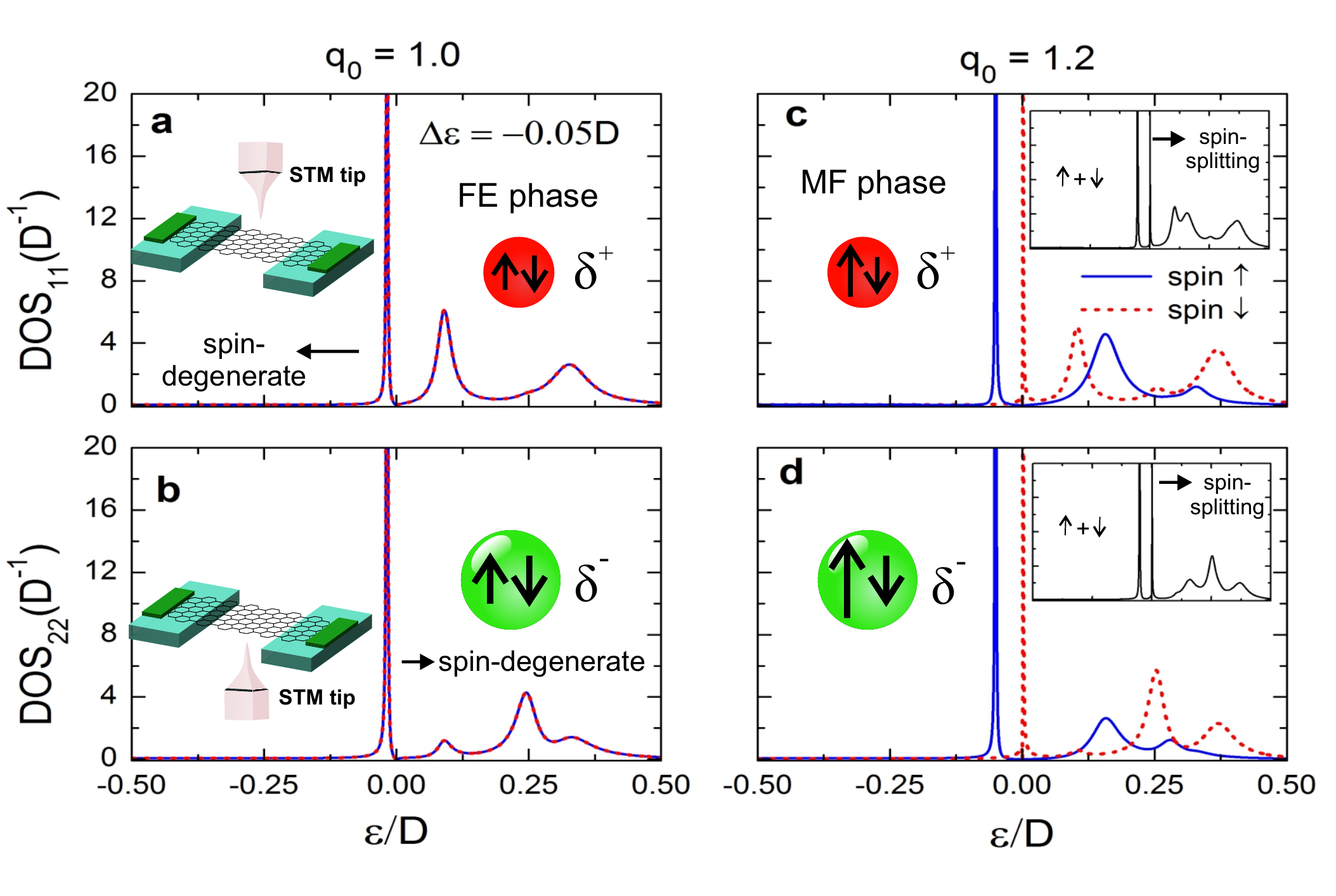}
\caption{\label{fig:Pic4}(Color online) (a)-(b) Spin-degenerate and charge
split DOSs of the adatoms, due to the finite dipole moment, representing
the FE phase, which can be probed by an STM tip. (c)-(d) In the MF
phase, the DOSs are simultaneously charge and spin split.}
\end{figure}

If one uses the detuning $\Delta\mathcal{E}$ between the two collinear
adatoms as driving parameter (which can be done e.g. by application
of a bias parallel to the plane of the system) and keeps the Fano
factor fixed, the system exhibits MF behavior in the finite range
$\Delta\varepsilon_{\text{{c}}1}<\Delta\varepsilon<\Delta\varepsilon_{\text{{c}2}}$
as it can be seen at Figs.\ref{fig:Pic2}(d,e). If the two adatoms
are equal, naturally, there is no dipole moment in the system and
their magnetizations are equal as observed in the point $\Delta\mathcal{E}=0,$
which is marked by the circle in the vertical dashed line in Fig.\ref{fig:Pic2}(d)
(FM phase). Notice that the dependence of $\delta^{-}-\delta^{+}$
on $\Delta\mathcal{E}$ shows linear trend as expected, with discontinuities
at the critical points $\Delta\varepsilon_{\text{{c}}1}$ and $\Delta\varepsilon_{\text{{c}}2}.$

Fig.\ref{fig:Pic3} represents the 3D plots showing magnetization
and dipole moment as a function of both $q_{0}$ and $\Delta\varepsilon$.
The presence of the QPT characterized by abrupt jumps of the dipole
moment and magnetization is clearly visible at these plots. The full
phase diagram of the system, which is the main result of the current
work is shown at Fig.\ref{fig:Pic3}(d).

As demonstrated in Ref.{[}\onlinecite{Shelykh}{]} by one of us,
one way to tune the slope of the Dirac cones is to couple electronic
states in free-standing graphene to linear polarized dressing light
field. To detect the FE phase, one can employ STM tip measurements of differential conductance for suspended
graphene\cite{Dresselhaus} (insets of Figs.\ref{fig:Pic4}(a,b))
which can probe $\text{DOS}_{ll}^{\uparrow}+\text{DOS}_{ll}^{\downarrow}.$
The FE feature is then revealed, as ensured by Eq.(\ref{eq:occupation}),
just by determining the areas under the curves of $\text{DOS}_{22}^{\uparrow}+\text{DOS}_{22}^{\downarrow}$
and $\text{DOS}_{11}^{\uparrow}+\text{DOS}_{11}^{\downarrow},$ which
give respectively distinct charge accumulations $\delta^{-}=<n_{2\uparrow}>+<n_{2\downarrow}>$
and $\delta^{+}=<n_{1\uparrow}>+<n_{1\downarrow}>$ that characterize
the local ordering parameter $\delta^{-}-\delta^{+}\neq0$,
since we can notice from panels (a,b) different areas. We should pay
particular attention that in the FE phase, a single pronounced spin-degenerate
peak nearby the Dirac point is also a hallmark of such a phase.

A sudden spin-splitting (panels (c,d) and insets with $q_{0}=1.2$),
due to the QPT, of the already charge split resonant states (panels
(a,b) with $q_{0}=1$) is verified in the MF phase. Such a splitting
can be detectable just by employing an unpolarized STM tip, which
reveals in $\text{DOS}_{ll}^{\uparrow}+\text{DOS}_{ll}^{\downarrow},$
a pair of peaks close to the Dirac point emerging instead of the only
single verified in FE phase, here appearing depicted in the insets
of panels (c,d) of the same figure and being very similar to the result
observed experimentally in Ref.{[}\onlinecite{STM}{]}, due to local
magnetic moments of adatoms. It means that besides the feature of
$\delta^{-}-\delta^{+}\neq0$ as aftermath of distinct areas under
the curves of $\text{DOS}_{22}^{\uparrow}+\text{DOS}_{22}^{\downarrow}$
and $\text{DOS}_{11}^{\uparrow}+\text{DOS}_{11}^{\downarrow}$ (panels
(c,d)), the pair of peaks in the neighborhood of the Dirac point (insets
of these panels) constitutes the capital fingerprint for confirming
the MF phase. Thus the presence of the FE ordering is revealed by
the charge split peaks in the DOSs as shown at Figs.\ref{fig:Pic4}(a,b),
while the transition into the MF phase is accompanied by the emergence
of the additional spin splitting of the peaks as shown at Figs.\ref{fig:Pic4}(c,d).
Thereby, we consider these features as the \textit{smoking-gun} of
the QPT transition from the FE phase to MF, which experimentalists
can pursuit for.

\section{Conclusions}

In summary, we have shown that graphene with collinear pair of magnetic
adatoms can be driven into a MF phase via a QPT by changing the slope
of the Dirac cones. To detect such a QPT, we claim that the proper
tool to this aim is the STM experiment reported in Ref.{[}\onlinecite{STM}{]}.

\section{Acknowledgments}

This work was supported by CNPq, CAPES, 2015/23539-8 S{ã}o Paulo
Research Foundation (FAPESP) and FP7 IRSES project QOCaN. I.A.S acknowledges
the support from Rannis project BOFEHYSS, Horizon2020 RISE project
CoExAn and 5-100 program of Russian Federal Government.


\begin{thebibliography}{10}
\bibitem{Khomskii}{Daniel Khomskii, Physics \textbf{2}, 20 (2009).}

\bibitem{Eerenstein}{W. Eerenstein, N. D. Mathur, and J. F. Scott,
Nature \textbf{442}, 759 (2006).}

\bibitem{Review1}Y. Tokura, S. Seki, and N. Nagaosa, Rep. Prog. Phys.
\textbf{77}, 076501 (2014).

\bibitem{Review2}R. Ramesh and N. A. Spaldin, Nature {Mater.} \textbf{6},
21 (2007).

\bibitem{Kimura}{T. Kimura, T. Goto, H. Shintani, K. Ishizaka, T.
Arima, and Y. Tokura, Nature \textbf{426}, 55 (2003).}

\bibitem{Hur}{N. Hur, S. Park, P. A. Sharma, J. S. Ahn, S. Guha,
and S-W. Cheong, Nature \textbf{429}, 392 (2004).}

\bibitem{Cheong}{S.-W. Cheong and M. Mostovoy, Nature Mater. \textbf{6},
13 (2007).}

\bibitem{Fiebig}M. Fiebig, J. Phys. D, \textbf{38}, R123 (2005).

\bibitem{MexicanH}L. Seixas, A. S. Rodin, A. Carvalho, and A. H.
Castro Neto, Phys. Rev. Lett. \textbf{116}, 206803 (2016).

\bibitem{FeTe}{M. Pregelj, A. Zorko, O. Zaharko, Z. Kutnjak, M.
Jagodic, Z. Jaglicic, H. Berger, M. de Souza, C. Balz, M. Lang, and
D. Arcon, Phys. Rev. B \textbf{82}, 144438 (2010).}

\bibitem{FSalts}G. Giovannetti, R. Nourafkan, G. Kotliar, and M.
Capone, Phys. Rev. B \textbf{91}, 125130 (2015).

\bibitem{Mariano}P. Lunkenheimer, J. Müller, S. Krohns, F. Schrettle,
A. Loidl, B. Hartmann, R. Rommel, M. de Souza, C. Hotta, J. A. Schlueter,
and M. Lang, Nature {Mater.} \textbf{11}, 755 (2012).

\bibitem{Dresselhaus}V. Meunier, A. G. Souza Filho, E. B. Barros,
and M. S. Dresselhaus, Rev. Mod. Phys. \textbf{88}, 025055 (2016).

\bibitem{CNeto}A. H. Castro Neto, F. Guinea, N. M. R. Peres, K. S.
Novoselov, and A. K. Geim, Rev. Mod. Phys. \textbf{81}, 109 (2009).

\bibitem{BUchoa}B. Uchoa, L. Yang, S.-W.Tsai, N. M. R. Peres, and
A. H. Castro Neto, New J. Phys. \textbf{16}, 013045 (2014).

\bibitem{Anderson}P. W. Anderson, Phys. Rev. \textbf{124}, 41 (1961).

\bibitem{Fano1}A. C. Seridonio, M. Yoshida, and L. N. Oliveira, Europhys.
Lett. \textbf{86}, 67006 (2009).

\bibitem{Overlaps}Z.-G. Zhu, K.-H. Ding, and J. Berakdar, Europhys.
Lett. \textbf{90}, 67001 (2010).

\bibitem{STM}H. G.-Herrero, J. M. G.-Rodríguez, P. Mallet, M. Moaied,
J. J. Palacios, C. Salgado, M. M. Ugeda, J.-Y. Veuillen, F. Yndurain,
and I. Brihuega, Science \textbf{352}, 437 (2016).

\bibitem{Hubbard}J. Hubbard, Proc. R. Soc. Lond. A, \textbf{281},
401 (1964).

\bibitem{Tfinite}P. Gegenwart, Q. Si, and F. Steglich, Nature Physics
\textbf{4}, 186 (2008).

\bibitem{book}R. Resta and D. Vanderbilt, Topics in Applied Physics,
\textbf{105}, 31, (2007).

\bibitem{Mariano1}S. R. Hassan, A. Georges, and H. R. Krishnamurthy,
Phys. Rev. Lett. \textbf{94}, 036402 (2005).

\bibitem{Mariano2}M. de Souza, A. Brühl, Ch. Strack, B. Wolf, D.
Schweitzer, and M. Lang, Phys. Rev. Lett. \textbf{99}, 037003 (2007).

\bibitem{Mariano3}M. de Souza and L. Bartosch, J. Phys. Cond.: Matter,
\textbf{27}, 053203 (2015).

\bibitem{Mariano4}M. de Souza, P.F.-Leylekian, A. Moradpour, J.-P.
Pouget, and M. Lang, Phys. Rev. Lett. \textbf{101}, 216403 (2008).

\bibitem{Mariano5}M. de Souza and J.-P. Pouget, J. Phys. Cond.: Matter,
\textbf{25}, 343201 (2013).

\bibitem{LH} L.H. Guessi,  R. S. Machado, Y. Marques, L. S. Ricco, K. Kristinsson, M. Yoshida, I. A. Shelykh, M. de Souza,
and A. C. Seridonio, Phys. Rev. B \textbf{92}, 045409 (2015).

\bibitem{Shelykh}K. Kristinsson, O. V. Kibis, S. Morina, and I. A.
Shelykh, Sci. Rep. \textbf{6}, 20082 (2016).\end{thebibliography}
\end{document}